# Photo- and thermally-induced huge layer decoupling in twisted bilayer $WSe_2$

A. Nakamura[1,2], Y. Chiashi[2,†], T. Shimojima[1,‡], Y. Tanaka[2], S. Akatsuka[2], M. Sakano[2,§], S. Masubuchi[3], T. Machida[3], K. Watanabe[4], T. Taniguchi[5], and K. Ishizaka[1,2,*]

[1]RIKEN Center for Emergent Matter Science, Wako, Saitama 351-0198, Japan.
[2]Quantum-Phase Electronics Center and Department of Applied Physics, The University of Tokyo, Hongo, Tokyo 113-8656, Japan.
[3]Institute of Industrial Science, The University of Tokyo, Meguro, Tokyo 153-8505, Japan.
[4]Research Center for Electronic and Optical Materials, National Institute for Materials Science, 1-1 Namiki, Tsukuba 305-0044, Japan
[5]Research Center for Materials Nanoarchitectonics, National Institute for Materials Science, 1-1 Namiki, Tsukuba 305-0044, Japan

**Present address:**
[†]National Institute of Advanced Industrial Science and Technology, Tsukuba, Ibaraki, 305-8568, Japan
[‡]Department of Physics, Nagoya University, Furo-cho, Aichi 464-8601, Japan
[§]Graduate School of Informatics and Engineering, The University of Electro-Communications, Chofu-gaoka, Tokyo 182-8585, Japan
Email: ishizaka@ap.t.u-tokyo.ac.jp

## Abstract

Twisted bilayer systems host a wealth of emergent phenomena, such as flat-band superconductivity, ferromagnetism, and ferroelectricity, arising from moiré superlattices and unconventional interlayer coupling. Despite their central role, direct and quantitative access to the three-dimensional atomic arrangement in these systems has remained elusive due to their nanoscale dimensions. Here, we introduce an automated dark-field electron tomography technique that enables quantitative three-dimensional structural analysis of atomically thin materials with sub-ångström precision. By applying this method to twisted bilayer $WSe_2$, we precisely visualize the twist-angle-dependent structural relaxation appearing as the AB/BA stacking domains separated by 10-20 nm domain walls. In the marginally twisted region ($\theta \leq 0.1°$), we uncover a significant expansion of the interlayer spacing compared to the bulk configuration, exceeding 0.1 Å, along with a remarkable temperature-driven interlayer decoupling. Ultrafast measurements further reveal optically induced interlayer separation of ~0.2 Å on the picosecond timescale, attributed to transient exciton formation. These findings not only establish a powerful approach for visualizing hidden out-of-plane structures in atomically thin micro-flake materials, but also uncover the intrinsic fragility and dynamical tunability of interlayer coupling in moiré-engineered 2-dimensional materials.

Twisted bilayers of van der Waals materials have opened new frontiers in condensed matter physics, offering a versatile platform for realizing strongly correlated, topological, and ferroelectric phases. When two atomically thin crystals are stacked with a slight rotational mismatch, long-wavelength moiré superlattices emerge, reshaping the band structure and vibrational modes through periodic modulation of the local atomic registry. These engineered systems exhibit an extraordinary array of phenomena, including flat-band superconductivity [1–3], gate-tunable ferromagnetism [4], and polar textures akin to ferroelectricity [5–7]. While graphene-based moiré systems have led this revolution, semiconducting transition metal dichalcogenides (TMDs) provide an equally rich landscape, where twist angle, stacking sequence, and layer composition can be precisely controlled to manipulate superconductivity [8,9], correlated insulating states [10–12], moiré excitons [13], phonon modes [14], and charge-density waves [15]. As such, twisted bilayer TMDs have rapidly become a model system for probing and designing emergent quantum states.

These phenomena originate from the three-dimensional (3D) atomic arrangement created when two atomically thin layers are twisted to form a moiré superlattice. In twisted bilayers, a small twist angle reorganizes the local atomic registry both in-plane and out-of-plane, giving rise to reconstructed AB/BA domains, domain walls, and a spatially modulated interlayer geometry across the moiré unit cell [16,17], as shown in Fig. 1(a,b). This reconstructed 3D structure strongly modifies the interlayer coupling, which in turn governs a wide range of emergent properties: interlayer hybridization underpins the formation of flat bands [1,2,10], out-of-plane polarization couples with in-plane shear to enable sliding ferroelectricity [5–7], and excitonic properties can be tuned by controlling the interlayer distance to favor intralayer or interlayer excitons [18]. Therefore, understanding moiré materials requires quantitative knowledge of both in-plane and out-of-plane structural degrees of freedom. However, these two aspects have so far been accessed in a highly unbalanced manner: in-plane structures have been extensively imaged using piezoresponse force microscopy, scanning tunneling microscopy (STM), and transmission electron microscopy (TEM), whereas quantitative measurements of out-of-plane (stacking-direction) structure have proven exceptionally challenging. Atomic force microscopy and theoretical modeling suggest interlayer spacing variations of 0.1–0.8 Å across the moiré superlattice [17,19–24], but three-dimensional reconstruction of the interlayer geometry remains elusive, posing a significant barrier to understanding and engineering moiré-driven phenomena.

In principle, diffraction measurements provide a route to extracting 3D structural information from two-dimensional (2D) materials, just as in conventional crystals. As an illustrative example, Fig. 1(c) shows a dark-field TEM image of the 0 2 diffraction from twisted bilayer $WSe_2$, revealing triangular AB/BA domains via intensity contrast. This contrast arises from the scattering-vector [$\boldsymbol{q} = (q_x, q_y, q_z)$] dependence of the diffraction intensity $S(\boldsymbol{q})$, which differs between AB and BA stacking due to their opposing rod-like scattering features along $q_z$. The $q_z$-resolved intensity encodes structural parameters such as the selenium height ($h$) and interlayer spacing ($d$), offering a pathway to quantify the out-of-plane structure from tilt-dependent diffraction patterns. However, conventional x-ray diffraction is not viable for atomically thin micro-flake samples. While positron diffraction has succeeded in millimeter-scale bilayer graphene [25], its applicability remains limited for micro-flakes as well. In contrast, TEM offers a compelling solution. As well as the sufficient spatial resolution that can access the AB/BA stacking domain structure, recent advances in automated 3D electron diffraction tomography enable high-precision refinement of 3D structural models [26,27] based on the diffraction intensity. Similar principles can be extended to atomically thin materials using tilt-dependent dark-field imaging [28], and further integrated with ultrafast TEM to resolve structural dynamics on picosecond timescales. These emerging techniques provide a powerful foundation for probing the elusive 3D crystal structure and its dynamics in 2D materials.

In this study, we introduce a method we term "automated dark-field tomography", which enables quantitative reconstruction of the 3D structure of twisted bilayer systems from tilt-series dark-field TEM images. Using this approach, we first reveal the twist-angle-dependent structural relaxation, showing the evolution from fully-relaxed AB/BA domains in marginally twisted regions to incompletely relaxed structures at larger twist angles. We then directly determine the interlayer spacing and chalcogen atom height in both twisted and naturally stacked bilayer $WSe_2$, revealing that both are significantly expanded compared to the bulk 2H-$WSe_2$ phase. Strikingly, we find that fully relaxed AB/BA domains of twisted bilayer $WSe_2$ exhibit sub-ångström interlayer decoupling upon heating, an effect absent in naturally stacked counterparts. Moreover, by integrating ultrafast TEM techniques [29,30], we uncover a light-induced, nonthermal expansion of the interlayer spacing on the picosecond timescale, with magnitudes exceeding those in bulk materials. These results establish a powerful new methodology for structural analysis of 2D materials and reveal the exceptional fragility and tunability of interlayer coupling in twisted bilayers, driven by both thermal and optical stimuli.

## Automated dark-field tomography in bilayer $WSe_2$

To begin with, we discuss the relationship between the dark-field image and the crystal structure of twisted- and natural-bilayer $WSe_2$. Figure 1(c) shows 0 2 dark-field images of the 0.10° twisted-bilayer $WSe_2$ at the sample tilt $\alpha = 10°$, 18°, and 24° [see Fig. 1(d) and Sec. 1 in Supplemental Materials]. Characteristic triangular patterns corresponding to the twist angle of 0.10° are observed in the dark-field images, where the dark/bright contrast changes as $\alpha$ is varied. This contrast arises from the local stacking configurations realized by the reconstruction of the moiré superlattice in small-angle twisted-bilayer TMDs [31], which can be understood as follows. Figures 1(e,f) show the crystal structure and kinematically simulated dark-field intensity profile $I_{02(0\bar{2})}(\boldsymbol{q})$ of AB and BA stacking bilayer $WSe_2$ as a function of $\boldsymbol{q}$ (see Sec. 1 in Supplemental Materials). Finite-size along the out-of-plane ($z$) direction in the twisted-bilayer system results in the rod-like $I_{02(0\bar{2})}(\boldsymbol{q})$ along $q_z$ direction. The dashed white line denotes the cross-section of the Ewald sphere, and the angle $\phi$ monotonically changes when the sample tilt $\alpha$ is changed. Therefore, the $q_z$ dependence of $I_{02(0\bar{2})}(\boldsymbol{q})$ can be obtained via the α-dependent dark-field intensities. The triangular contrast in Fig. 1(c) corresponds to the difference of $I_{02}(q_z)$ in local stacking domains. As such dark-field intensity is strongly dependent on the local crystal structure, it can be used for real-space visualization of the stacking configurations including boundary in principle. Furthermore, the analytical expression of $I_{02}(q_z)$ for fully relaxed AB/BA stacking region contains the full atomic coordinates of $WSe_2$, thus we can quantitatively determine the 3D crystal structure parameters from automatically collected $\alpha$-dependent dark-field image intensity $I_{02}(\boldsymbol{q})$ (see Sec. 1 in Supplemental Materials). This method, which we refer to as automated dark-field tomography, can be used for atomically thin flake samples in general. Natural-bilayer $2H$-$WSe_2$ also shows rod-like $I_{02}(\boldsymbol{q})$ [Figs. 1(g,h)] with $q_z$ dependence significantly different from AB/BA stacking. Thus the crystal structure parameters of natural-bilayer 2H-$WSe_2$ can also be determined from automated dark-field tomography.

**Twist-angle dependent relaxation of crystal structure**

Here, we investigate how the crystal structure relaxation evolves as a function of twist angle $\theta$ using automated dark-field tomography. Although the target twist angle during sample preparation was 0.3°, we observe a range of domain sizes at different positions within the same flake due to the extrinsic inhomogeneity. Fig. 2(a) shows the dark-field intensity $I_{02}(\alpha)$ curves taken within the AB/BA stacking domains and at the domain wall under marginally twisted ($\theta$ <0.05°) conditions. As described above, $I_{02}(\alpha)$ exhibits distinct peak structures depending on the local stacking configurations. As a result, the dark-field images at $\alpha = 24°$ [Fig. 2(b-d)] visualize the stacking domains. To qualitatively evaluate the local stacking configurations, we focused on the angular shift of the peak in $I_{02}(\alpha)$ between 0° and 10°. This peak in Fig. 2(a) monotonically changes among the AB domain, BA domain, and domain wall regions, suggesting that its relative position is sensitive to the local stacking configuration across the domain wall. Based on this observation, we use the relative spatial variation of the center of mass (CoM) between 0° and 10° [Fig. 2(e-g)] to visualize shifts in the $I_{02}(\alpha)$ peak position associated with local stacking configurations. It should be noted that the absolute CoM value appears as "angle" but cannot be directly assigned to any structural parameter, as the peak is very broad and the CoM depends on the selected angular window. CoM in Fig. 2(e,f) clearly visualizes the domain walls (appearing as yellow in the CoM colorscale) as well as the fully relaxed AB/BA domain regions (brown/light green). Here we note that the present spatial resolution is better than the pixel size (6.32 nm). The CoM profile in Fig. 2(h,i) exhibits a clear step-like change between the AB and BA domains, reflecting the presence of a well-defined domain wall between fully relaxed AB/BA domains. The full width of this step is approximately 15 nm, which is comparable to that reported for bilayer graphene (~10 nm) [32]. The CoM becomes nearly flat within about 10 nm away from the domain wall suggesting that the AB/BA stacking relaxation is essentially complete over most of the domain area. At a larger twist angle of $\theta$ =0.27° in Fig. 2(g), the contrast is lower than that at smaller twist angles, implying incomplete relaxation in the AB/BA stacking regions. The CoM profile in Fig. 2(j) no longer shows a step-like feature separated by flat AB/BA regions, but instead exhibits a more gradual variation across the moiré period. This indicates that the stacking configuration is continuously modulated over the entire region, rather than being fully relaxed into AB/BA domains separated by narrow domain walls. Notably, this incomplete relaxation occurs even though the moiré period is still about five times larger than the domain-wall width observed at smaller twist angles.

The present results reveal a distinct feature of domain walls in twisted-bilayer systems. Previous studies have shown that AB/BA stacking reconstruction becomes significant below about 1° in both twisted-bilayer graphene and transition metal dichalcogenides [16,17], which is consistent with the clear AB/BA domain contrast of the dark-field images in Fig. 2(b-d). Nonetheless, the present results show that deviation from the fully relaxed AB/BA stacking configuration already exists across nearly the entire region even at a relatively small angle ($\theta$ =0.27°) due to the intrinsic strain caused by twisted structure, a point that has not been fully discussed in previous studies. In contrast, we found nearly homogeneous AB/BA stacking when $\theta \leq 0.10°$, probably because the domain size is much larger than the characteristic domain-wall width. These results indicate that, at twist angles below 0.10°, most of the AB/BA domain region is nearly fully relaxed into AB/BA stacking configurations while the local structure remains incompletely relaxed over nearly the entire domain area at θ = 0.27°.

**Crystal structure refinement of twisted- and natural-bilayer $WSe_2$**

Here, we experimentally determine the interlayer distance of twisted- and natural-bilayer $WSe_2$ using automated dark-field tomography. We focus on the twist angle $\theta \leq 0.10°$ because our structural refinement is based on the assumption that the local crystal structure is fully relaxed to AB/BA stacking. Figure 3(a) shows the dark-field intensity $I_{02}(\alpha)$ curves taken within the AB/BA stacking domains of $\theta = 0.10°$, as marked by the blue/red rectangles in the inset image. We fitted them by the squared structure factor $|S_{02}(\alpha, d_{\mathrm{TBL}}, h_{\mathrm{TBL}})|^2$ to determine $d_{\mathrm{TBL}}$ and $h_{\mathrm{TBL}}$ [see Sec. 1 in Supplemental Materials for details]. We assumed that the selenium height $h_{\mathrm{TBL}}$ is identical for all selenium atoms for simplicity. The solid gray curves in Fig. 3(a) are the fitting results that most well reproduce the experimental data. Here we obtained the structural parameters as $d_{\mathrm{TBL}} = 6.64 \pm 0.10$ Å and $h_{\mathrm{TBL}} = 1.76 \pm 0.10$ Å. We also measured the natural-bilayer $2H$-$WSe_2$, as shown in the inset in Fig. 3(b). Similarly to the twisted-bilayer sample, $I_{02}(\alpha)$ of the bilayer $2H$-$WSe_2$ was taken at the blue rectangle region (inset) and fitted by the gray solid curve. $d_{2H}$ and $h_{2H}$ are obtained as 6.63 $\pm$ 0.10 Å and 1.77 $\pm$ 0.10 Å, respectively.

The results of crystal structure refinements for twisted- and natural- bilayer $WSe_2$ are summarized in Fig. 3(c), along with the literature values for bulk 2H-$WSe_2$ [33]. The results of structural relaxation from density-functional theory (DFT) calculations are also plotted by filled markers (see methods). Compared with bulk 2H-$WSe_2$, both twisted- and natural-bilayer $WSe_2$ show a significantly large interlayer distance *d*, although the selenium height *h* is seemingly scattered within the experimental error. The larger interlayer distance observed in both bilayer systems suggests a reduced interlayer coupling, which should significantly affect their electronic and optical properties, as discussed later. The DFT calculation results (filled markers) also show the significant difference between the interlayer distance of bulk and bilayer samples, well reproducing the experimental results. This result gives the validation of the DFT calculations in twisted-bilayer systems, at least in the relaxed AB/BA regions.

**Thermally-induced sub-Å interlayer decoupling**

We further perform the temperature-dependent automated dark-field tomography of natural- and twisted-bilayer $WSe_2$, and reveal the thermally induced sub-Å interlayer decoupling in the twisted-bilayer sample. Fig. 3(d) shows the temperature-dependent $I_{02}(\alpha)$ curves for $24° < \alpha < 35°$, taken at the BA stacking region of $\theta = 0.10°$. At 400 K, the peak position is slightly shifted towards the lower $\alpha$ direction compared to 300 K. On the other hand, $I_{02}(\alpha)$ for the natural-bilayer sample [Fig. 3(e)] does not show any significant shift. Similar behaviors were also observed in the AB stacking counterpart, other $\alpha$ region, and marginally twisted regions ($\theta < 0.05°$) [Sec. 2 in Supplemental Materials]. Correspondingly, the fitting results of the twisted-bilayer sample show the change in the interlayer distance $\Delta d_{\mathrm{TBL}} = d_{\mathrm{TBL}}(T = 400\ \mathrm{K}) - d_{\mathrm{TBL}}(T = 300\ \mathrm{K})$ of 0.08 Å [Fig. 3(f)], whereas no significant change is observed in the natural-bilayer sample [Fig. 3(g)]. The estimated linear thermal expansion coefficient along $z$ for the twisted-bilayer sample is $\beta = 8.0 \times 10^{-5}\ \mathrm{K}^{-1}$, which is much larger than those of the natural-bilayer sample ($< 2.0 \times 10^{-5}\ \mathrm{K}^{-1}$) and the bulk $2H$-$WSe_2$ sample [34] ($1.0 \times 10^{-5}\ \mathrm{K}^{-1}$) as summarized in Fig. 3(h). This value is comparable to polymers such as polyvinyl chloride, showing larger thermal expansion compared to inorganic solids [35]. These results reveal an anomalously large thermally induced sub-Å interlayer decoupling in fully relaxed AB/BA domains of twisted bilayer $WSe_2$, far exceeding that of the natural 2H bilayer. However, since the present comparison is effectively made between a 3R-like twisted region and a natural 2H bilayer, the current experiment alone cannot determine whether this behavior should be attributed to the 3R-stack-like local stacking itself or to the twisted bilayer prepared by tear-and-stack method. Direct comparison with natural 3R and 2H bilayers, for example in $MoS_2$, will be an important direction for future work because no bulk 3R-$WSe_2$ crystal has been reported from which an equivalent reference bilayer could be prepared.

This anomalously large thermal expansion and sub-Å interlayer decoupling observed exclusively in the twisted-bilayer sample may originate from its unique bonding configuration associated with the catenary-like interlayer state. In twisted bilayers, the interlayer coupling is highly nonuniform across the moiré superlattice: theoretical studies [23,17,20] have predicted a substantial difference in the interlayer spacing between the AA stacking regions, where the layers are maximally separated at the vertices of the triangular pattern, and the AB/BA stacking regions, where the layers are most closely spaced, with a magnitude on the order of several tenths of an ångström. A recent STM study also reveals the out-of-plane atomic reconstruction depending on the stacking [36]. The present results, obtained from the structurally relaxed AB/BA regions, already indicate an unusually strong anharmonicity in the interlayer bonding potential of the twisted-bilayer system. Such enhanced anharmonicity may naturally lead to a large out-of-plane thermal response, far exceeding that of natural bilayers and bulk crystals. Although the current analysis can be applied to only fully relaxed AB/BA regions, further improvement of the structure model will enable us to access local crystal structure for larger twist angles, AA regions and the ridge-like domain boundaries of the moiré lattice. Probing these local structures is expected to provide decisive insights into how structural relaxation, anharmonicity, and thermally induced interlayer decoupling evolve with moiré geometry in the future.

### Non-thermal interlayer decoupling induced by pulsed light

We further investigate the photoinduced structural dynamics of twisted-bilayer $WSe_2$ (Sample #2) using ultrafast TEM [Fig. 4(a)]. Sample #2 is characterized in Sec. 3 in Supplemental Materials. Figure 4(b) shows the 0 2 dark-field image $I_{02}(t)$ of twisted-bilayer $WSe_2$ before photoexcitation, where triangular patterns are observed. Although these reconstructed AB/BA stacking domains in Fig. 4(b) were stable under equilibrium conditions for more than a month, they change abruptly and irreversibly when the sample is continuously photoexcited by a sequence of 25 kHz 2.4 eV light pulses for several days [see Fig. 4(c) and Sec. 3 in Supplemental Materials]. The AB/BA domains at the center of the image lose the ordered triangular shape, but after one week, the structural AB/BA domain shown in Fig. 4(c) was stable with a marginal twist angle of less than 0.05°.

Here, we investigate the crystal structure dynamics of such marginally twisted $WSe_2$ from the time-dependent change in dark-field image intensity. Figure 4(d) shows the 0 2 dark-field image of twisted-bilayer $WSe_2$ measured at $\alpha$ =24.1°. When the 2.4 eV light pulse of 0.33 mJ/cm$^2$ excites the sample, the dark-field-image intensity in the AB (BA) stacking region decreases (increases), as shown in Figs. 4(e,f). We found that triangular AB/BA domains also show a similar trend, as in Fig. S12 in Supplemental Materials, suggesting that the observed domain-dependent change in dark-field images is universal regardless of the domain shape. To quantitatively analyze the dark-field intensity, we subtracted the background and obtained the normalized dark-field intensity $\bar{I}_{02}(t)$ for AB/BA domains as shown in Fig. 4(g) (see Sec. 4 in Supplemental Materials for details). $\bar{I}_{02}(t)$ of the AB (BA) domain decreases (increases) immediately after the photoexcitation, within the time resolution 4 ps, and then relaxes back.

From the transient dark field intensity change, we further evaluate the time dependence of the interlayer distance $d_{\mathrm{TBL}}(t)$, revealing a sub-Å change in the interlayer distance. By comparing $I_{02}(t)$ and the simulated crystal structure factor $|S_{02}(\alpha, d_{\mathrm{TBL}}, h_{\mathrm{TBL}})|^2$ quantitatively, we can determine $d_{\mathrm{TBL}}(t)$. For BA stacking, the effect of $h_{\mathrm{TBL}}(t)$ on $I_{02}(t)$ is negligible, and we can only evaluate $d_{\mathrm{TBL}}(t)$ (see Sec. 5 in Supplemental Material). As shown in Fig. 5(a), the obtained $\Delta d_{\mathrm{TBL}}(t) = d_{\mathrm{TBL}}(t) - d_{\mathrm{TBL}}(t < 0)$ increases by 0.2 Å at *t* ~ 0 and subsequently relaxes. Regarding the AB stacking, on the other hand, $h_{\mathrm{TBL}}$ and $d_{\mathrm{TBL}}$ both contribute to $I_{02}(t)$ and they cannot be independently evaluated. Here we adopt $\Delta h_{\mathrm{TBL}}(t) = h_{\mathrm{TBL}}(t) - h_{\mathrm{TBL}}(t < 0) = 0$, which provides the solution of $\Delta d_{\mathrm{TBL}}(t)$ similar to the BA stacking (see Sec. 5 in Supplemental Material).

As shown in Fig. 5(a), in the relaxation process, $\Delta d_{\mathrm{TBL}}$ apparently shows the fast (about 50 ps) and slow (> 200 ps) relaxation components. To discuss the origin of these relaxation components, we evaluated the change in the transient lattice temperature $\mathit{\Delta} T$ from the Debye-Waller factor of 4 0 diffraction intensity, which is further normalized by that just after the photo-excitation ($\mathit{\Delta} T_{\mathrm{max}}$) (see Sec. 5 in Supplemental Materials). The $\mathit{\Delta} T/\mathit{\Delta} T_{\mathrm{max}}$ curve in Fig. 5(b) behaves differently from $\Delta d_{\mathrm{TBL}}$: an exponential decay curve with a single time constant of 200 ps can well describe $\mathit{\Delta} T/\mathit{\Delta} T_{\mathrm{max}}$ curve between 0 and 400 ps. These results indicate that the photoinduced change in the interlayer distance cannot be attributed to the change in the transient lattice temperature and should be non-thermal, at least in the fast (< 50 ps) timescale.

The appearance of fast and slow decay components in the interlayer distance is reminiscent of exciton dynamics observed in monolayer samples [37–40]. When the exciton density is sufficiently high in these systems, relatively fast (< 100 ps) decay of excitons due to the exciton-exciton annihilation process has been reported in addition to the slow (100 ps ~ few ns) recombination process. Such an exciton-exciton annihilation process is not observed in bulk and, therefore, is unique to atomically thin samples [39]. These previous studies indicate that the non-thermal interlayer decoupling observed in Fig. 5(a) may be attributed to the positive stress due to the exciton generation that is located at around 1.5-1.7 eV in twisted-bilayer $WSe_2$ [41]. The stress caused by the semiconductor carriers, including excitons, is usually described by deformation potential, which is related to the pressure-dependent electronic band structure [42]. Considering that the hydrostatic pressure significantly affects the interlayer coupling and bandgap in the bilayer system [43], the deformation potential mechanism caused by the exciton is a plausible scenario for the change in the interlayer distance. In addition, we also mention that the $\Delta d_{\mathrm{TBL}}$ at 400 ps [Fig. 5(a)] is not fully relaxed to the initial state while $\Delta T/\Delta T_{\max}$ is almost zero [Fig. 5(b)]. This indicates the relatively slow ($t$ > 400 ps) exciton recombination process, in addition to the fast exciton-exciton annihilation (< 50 ps) and thermal relaxation (~ 200 ps) as summarized in Fig. 5(c). To further verify the exciton-induced interlayer-decoupling scenario, future measurements with varying photon energy and fluence would be valuable. In particular, photon-energy dependence could help identify which excitonic states contribute to the interlayer decoupling.

The observed 0.2 Å change in interlayer distance is significantly larger as compared to previous studies on various bulk materials [43,44]. In bulk graphite [43], the magnitude of the change in interlayer distance is less than 0.01 Å even at a pump fluence of 11 mJ/cm$^2$, a value significantly different from the present result (0.2 Å at 0.33 mJ/cm$^2$). $\Delta d_{\mathrm{TBL}}$ = 0.2 Å is comparable to the 25-50% interlayer distance variation between AA and AB/BA stacking predicted by the theoretical calculations [23,17,20]. Although we do not observe any reversible change in the AB/BA stacking domain shape on the picosecond scale, frequent repetition of such a giant change in the interlayer distance may induce the deformation of the local twist angle and the resultant disordered domain structures, as seen in Fig. 4(b,c). Such a photoinduced domain reconstruction should be further investigated in future studies.

**Discussion: Impact of sub-Å interlayer decoupling**

The observed thermally and optically induced sub-Å interlayer decoupling will significantly impact various properties of twisted materials. The previous experiments reported that the properties of twisted-bilayer systems are significantly modified via a change in the interlayer distance, e.g., tuning of superconductivity [45], change in the indirect bandgap [20], formation of commensurate stacking [46], and modification of the electronic van Hove singularities [24]. In addition, many theoretical calculations [20,19,21,24,47,48] have revealed the band structure modification due to the change in the interlayer distance because it can directly affect the interlayer coupling. Based on the theoretical calculation in Ref. [20], $\Delta d_{\mathrm{TBL}}$ = 0.2 Å changes the indirect bandgap of $MoS_2$ by about 100 meV, although the direct bandgap at the K point is not significantly modified [20]. Such a large modification of the bandgap may induce direct-indirect bandgap transition in twisted-bilayer TMD semiconductors. Therefore, the thermally and optically induced sub-Å interlayer decoupling revealed by the present work should be essential in interpreting future research on twisted-bilayer systems. In particular, an interlayer decoupling effect caused by the excitons should play an essential role in the optical nature of the twisted-bilayer system.

**Conclusions**

In conclusion, we investigated the thermally and optically induced interlayer decoupling in twisted-bilayer $WSe_2$ using TEM. A newly developed automated dark-field tomography enables visualization of the local stacking configurations, including the boundary regions, and reveals that full relaxation into AB/BA stacking occurs only when the domain size is at least about one order larger than the domain-wall width. This technique further quantitatively determines the three-dimensional crystal structure of twisted- and natural-bilayer $WSe_2$, revealing that both twisted- and natural-bilayer samples show interlayer decoupling by 0.15 Å compared to bulk samples. Additional thermally-induced interlayer decoupling is also observed only in the twisted-bilayer sample, which is significantly larger than the bulk and natural-bilayer samples. Further investigation of picosecond crystal structure dynamics after photoexcitation revealed a non-thermal interlayer decoupling reaching 0.2 Å, probably due to the exciton generation. Such sub-Å interlayer decoupling in equilibrium and non-equilibrium states should significantly impact the electronic and optical properties of twisted-bilayer materials, and should be carefully considered in future studies. Application of the TEM-based methods presented in this study to a variety of materials/phenomena, such as sliding ferroelectricity [5–7], moiré enhanced charge-density wave [15], and light-induced ferromagnetism [49], is a promising direction in the future.

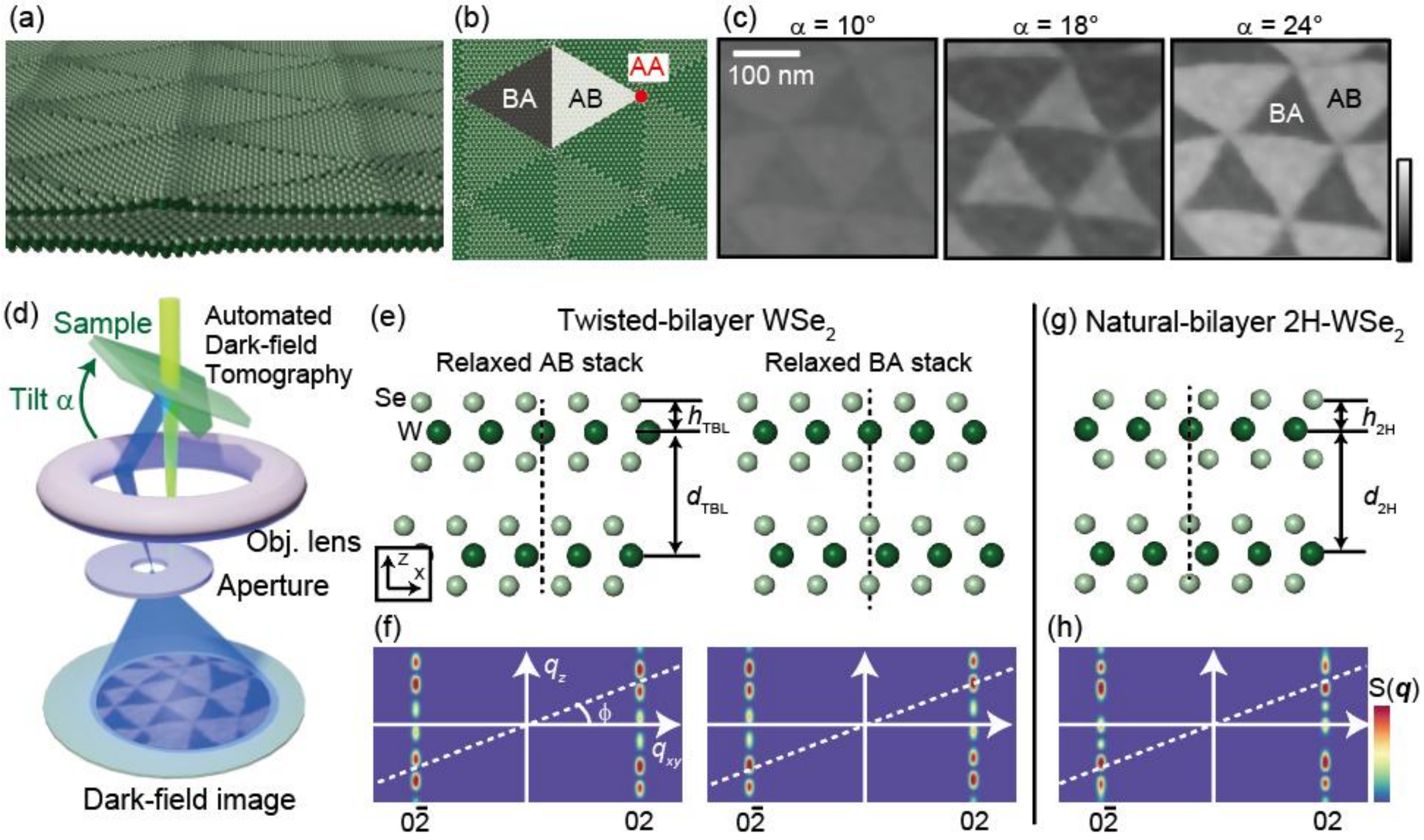

**Fig. 1.** (a) Schematic of superlattice relaxation in twisted-bilayer $WSe_2$. Dark- and light-green spheres denote tungsten and selenium atoms, respectively. For clarity, the spatial variation of interlayer distance due to the relaxation is emphasized. (b) Top view of the twisted-bilayer $WSe_2$. AA, AB, and BA stand for the AA, AB, and BA stacking regions of $WSe_2$ layers (c) 0 2 dark-field images of 0.1° twisted-bilayer $WSe_2$ at the sample tilt $\alpha = 10°$, 18°, and 24°, respectively. The triangular contrast reflects the AB/BA stacking domains. (d) Schematic of the automated dark-field tomography. Dark-field images of twisted-bilayer $WSe_2$ are recorded by changing the sample tilt α. (e) Side view of the twisted-bilayer $WSe_2$ at relaxed AB and BA stacking regions. $d_{\mathrm{TBL}}$ and $h_{\mathrm{TBL}}$ denote interlayer distance and selenium height. (f) The calculated diffraction intensity profile for AB, BA stacking bilayer $WSe_2$, which is squared crystal structure factor $|S(\boldsymbol{q})|^2$ described in Sec. 1 in Supplemental Materials. The dashed white line denotes the Ewald sphere corresponding to $\alpha$ =18°. (g) Side view of natural-bilayer $2H$-$WSe_2$. $d_{2\mathrm{H}}$ and $h_{2\mathrm{H}}$ denote interlayer distance and selenium height. (h) The calculated diffraction intensity profile for 2H stacking bilayer $WSe_2$.

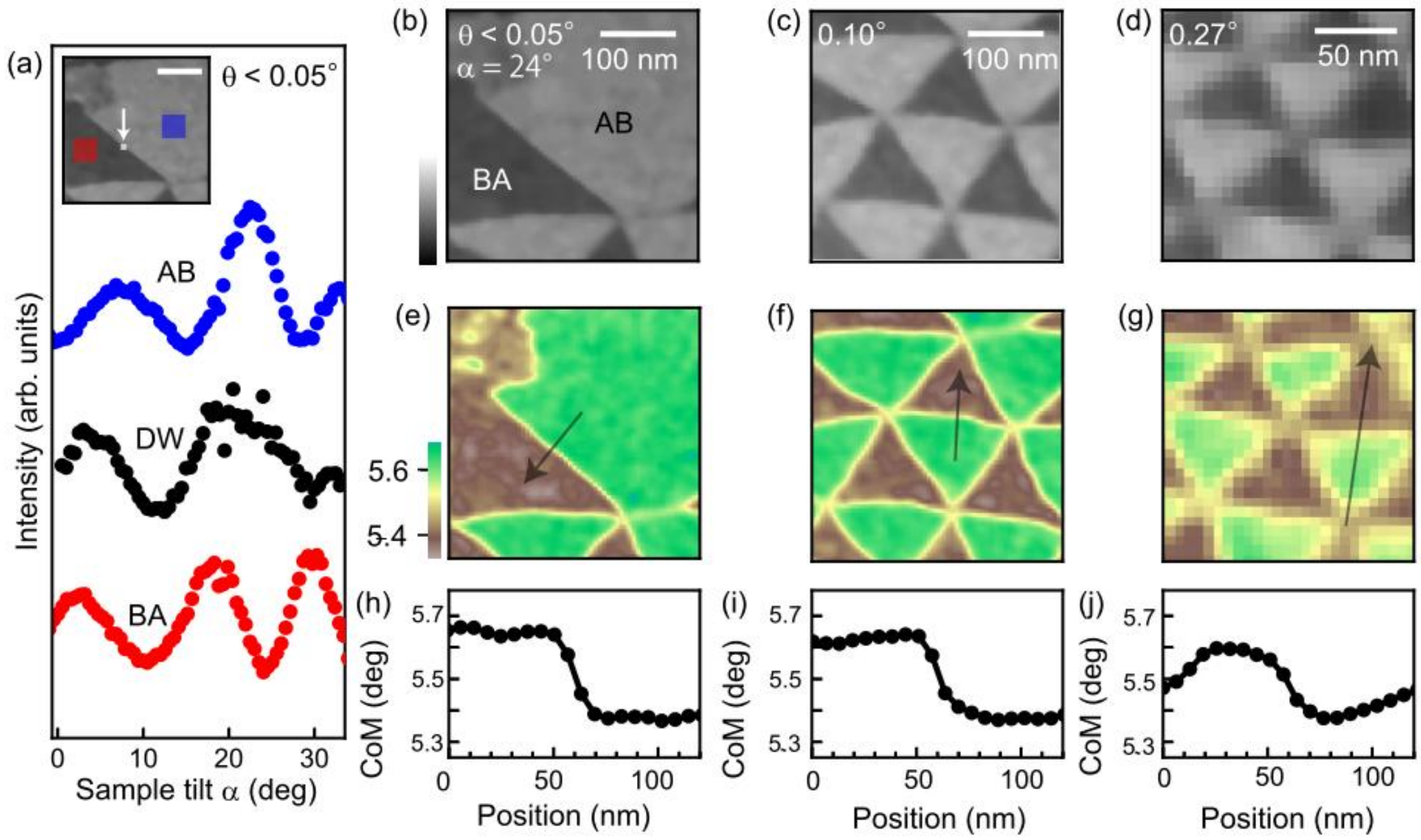

**Fig. 2.** (a) Automated dark-field tomography data of 0 2 diffraction $I_{02}(\alpha)$ for AB/BA stacking domains and domain wall (DW) in marginally-twisted-bilayer $WSe_2$ ($\theta$< 0.05°). The blue, red, and small white regions in the inset show where $I_{02}(\alpha)$ is obtained. The white region is also indicated by white arrow for clarity. (b-d) Twist-angle dependence of 0 2 dark-field images of twisted-bilayer $WSe_2$. (e-g) Twist-angle dependence of the center of mass (CoM) between 0° and 10°. (h-j) The profile of CoM along the black arrows in e-g.

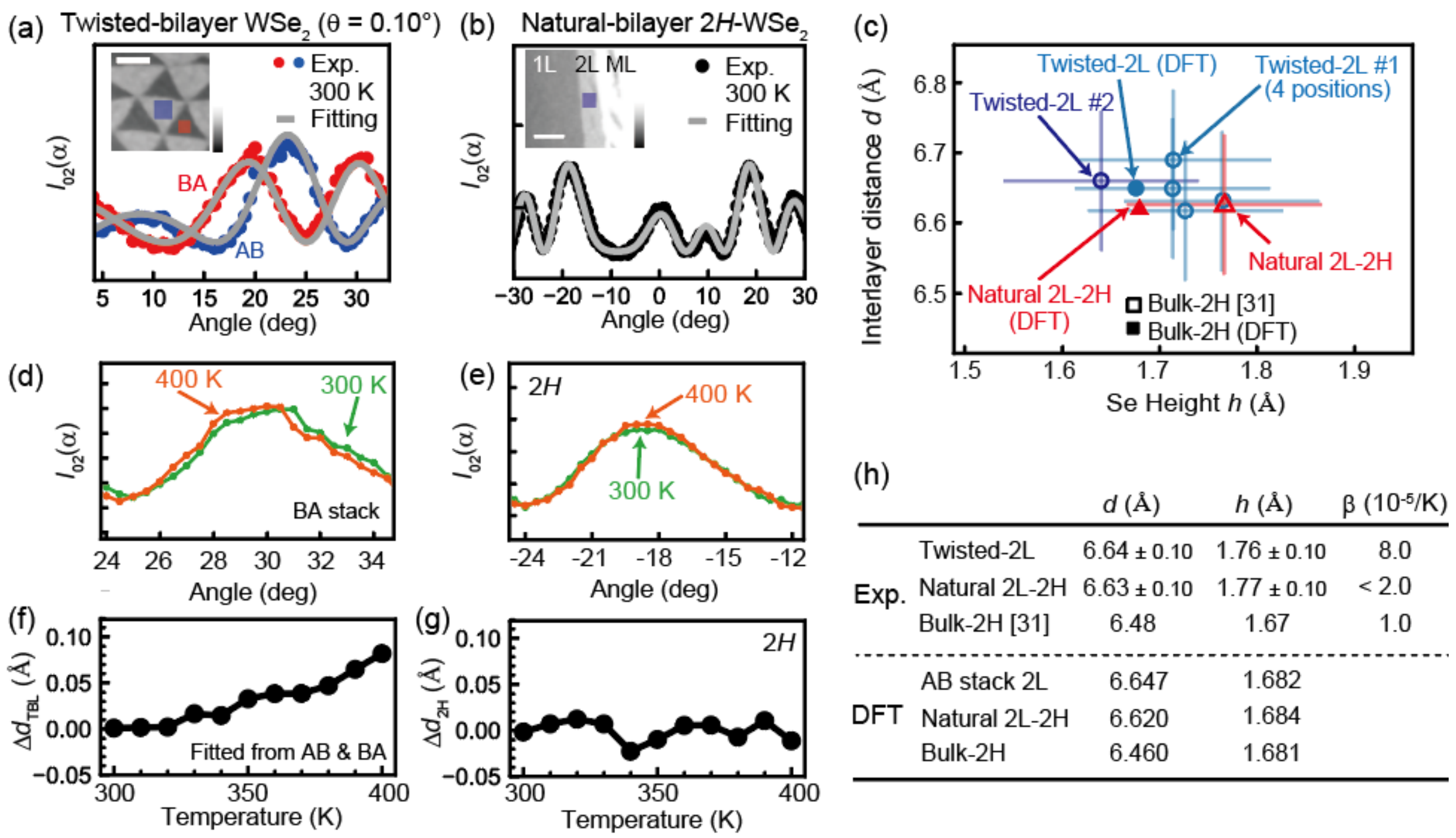

**Fig. 3.** (a) Automated dark-field tomography data of 0 2 diffraction $[I_{02}(\alpha)]$ of AB and BA stacking domains in twisted-bilayer $WSe_2$. The blue and red regions in the inset show where $I_{02}(\alpha)$ is obtained. The solid gray curves denote fitting results. The white scale bar in the inset denotes 100 nm. The vertical axes of a-d are on a linear scale. (b) $I_{02}(\alpha)$ of natural-bilayer $2H$-$WSe_2$. The blue rectangle in the inset shows where $I_{02}(\alpha)$ is obtained. In the inset, 1L, 2L, and ML indicate monolayer, bilayer, and multilayer $WSe_2$ obtained from the bulk $2H$-$WSe_2$ by exfoliation. The white scale bar in the inset denotes 500 nm. (c) Interlayer distance $(d_{\mathrm{TBL}}, d_{2H})$ and selenium height $(h_{\mathrm{TBL}}, h_{2H})$ obtained from automated dark-field tomography. Blue open circles denote the result of twisted-bilayer $WSe_2$, while the red open triangle is that of natural-bilayer $2H$-$WSe_2$. We performed five independent measurements on the twisted-bilayer $WSe_2$ by changing the sample (#1, #2) and position (4 positions on sample #1) in the sample. The black rectangle is the interlayer distance and selenium height of bulk $2H$-$WSe_2$ in the literature [33]. The filled markers denote the DFT calculation results (see methods). (d,e) Magnified $I_{02}(\alpha)$ curve measured at 300 K and 400 K. (f,g) Temperature dependence of the interlayer distance $d_{\mathrm{TBL}}$ and $d_{2H}$ obtained from the fitting (see Sec. 1 in Supplemental Materials). (h) The list of the interlayer distance $d$, selenium height $h$, and linear thermal expansion coefficient $\beta$ obtained from the automated dark-field tomography and density-functional calculations. For the twisted-bilayer $WSe_2$, we use the results obtained from a.

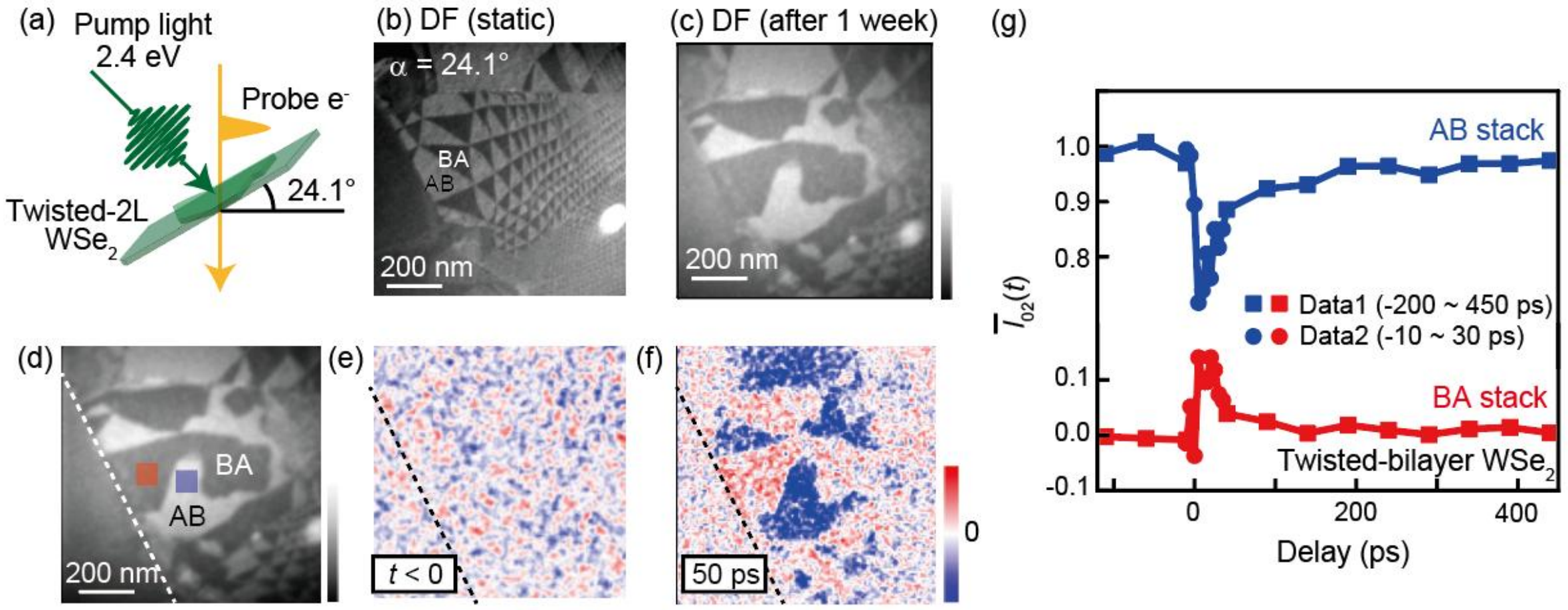

**Fig. 4.** (a) Schematic of ultrafast TEM measurement on twisted-bilayer $WSe_2$ (see methods for details). The sample tilt α was fixed to 24.1°. (b,c) 0 2 dark-field images of twisted-bilayer $WSe_2$ before and one week after continuous photoexcitation by a 2.4 eV light pulses. The bright and dark regions denote AB and BA stacking domains, respectively. (d) The 0 2 dark-field image of twisted-bilayer $WSe_2$ before photoexcitation ($t$ < 0). There is a monolayer region at the bottom-left region, denoted by the white dashed line. (e,f) Change in 0 2 dark-field images $I_{02}(t) - I_{02}(t < 0)$ before and after photoexcitation. (g) A background-subtracted normalized dark-field image intensity $\bar{I}_{02}(t)$ at the blue and red rectangles in d (see Sec. 4 in Supplemental Materials for details).

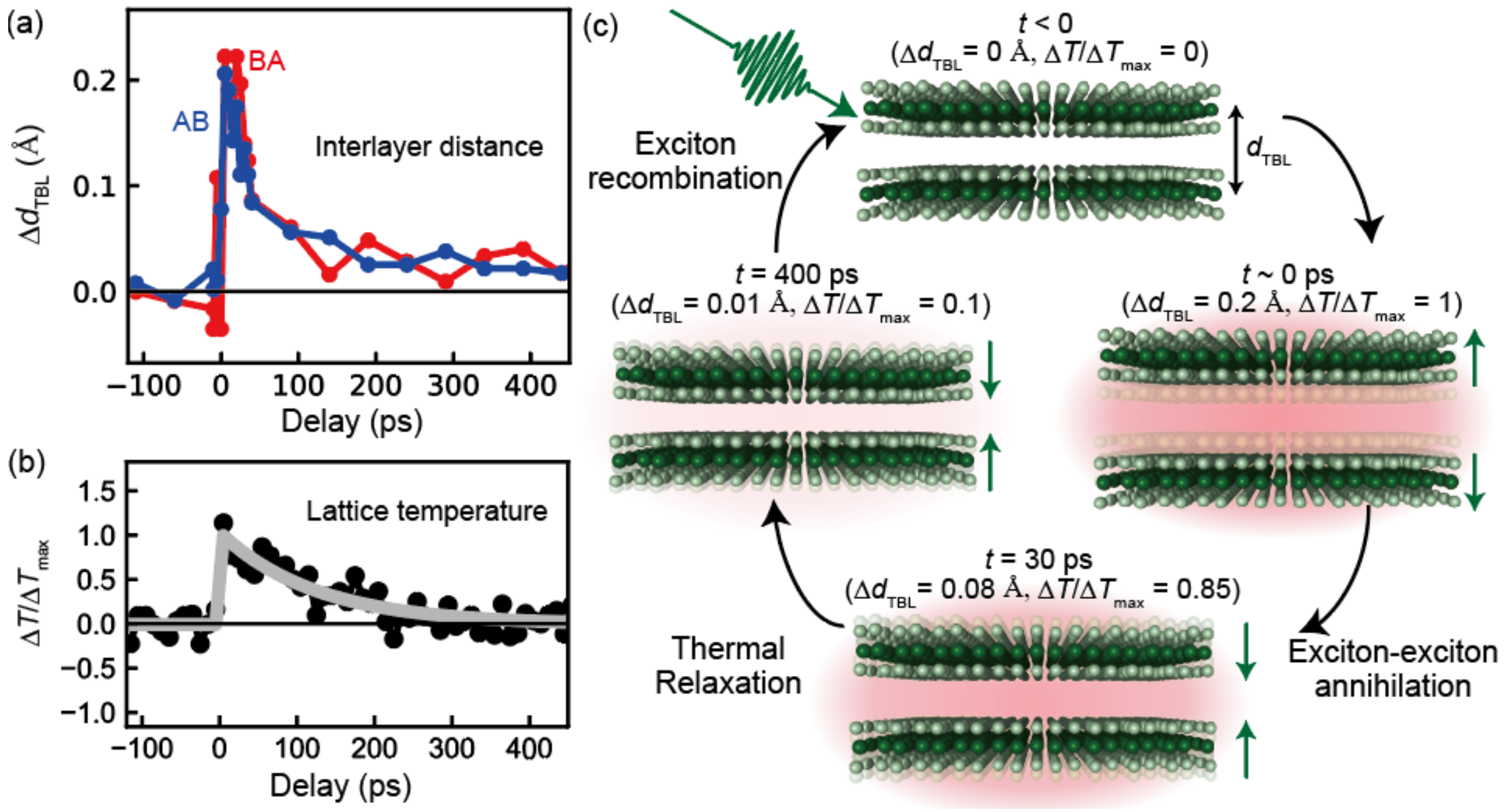

**Fig. 5.** (a) Time-dependence of the relative change in interlayer distance $\Delta d_{\mathrm{TBL}}$ in AB/BA stacking region estimated from the automated dark-field tomography results (see Sec. 4 in Supplemental Materials for details). (b) Time dependence of the change in lattice temperature $\Delta T$ normalized by that just after photoexcitation $\Delta T_{\mathrm{max}}$. The solid gray curve denotes the single exponential function with a time constant $\tau$ = 200 ps [see Sec. 5 in Supplemental Materials for details]. (c) Schematic of optically-induced interlayer decoupling in AB/BA stacking twisted-bilayer $WSe_2$. The interlayer distance $\Delta d_{\mathrm{TBL}}$ is changed by 0.2 Å by the photoexcitation.

# Methods

## Setup of (ultrafast) transmission electron microscopy

Transmission electron microscopy measurements were performed using a previously described system [29] based on Tecnai Femto (Thermo Fisher Scientific). The 150-μm condenser aperture was used for all measurements. We used a 20-μm objective aperture for bright-field and dark-field measurements unless otherwise denoted. A selected-area aperture of 200 μm was used for the diffraction measurements. For the automated dark-field tomography measurements, sample-tilt-dependent dark-field images are automatically collected with an angle step of 0.5°. For the time-resolved measurements in Fig. 4, we use 1030-nm light source PHAROS (Light Conversion) with the repetition rate of 25 kHz. We used frequency-doubled 514-nm light with a fluence of 330 $\mu J/cm^2$ as the pump light. The total time resolution was set at 4 ps for data acquisition to maintain a good signal-to-noise ratio throughout the entire measurement, although the best time resolution of our ultrafast transmission electron microscopy system was 750 fs. The beam diameter of the pump light (~160 μm) was set much larger than that of the sample (< 10 μm). The spatial resolution of the dark-field imaging was measured using a 0 2 dark-field image at bilayer-monolayer boundary region under the same experimental conditions as those used in Fig. 2. As shown in Fig. S13, the dark-field intensity is changed within one pixel of the image. This result indicates that the present measurements achieve a spatial resolution better than the pixel size (6.32 nm).

## Sample preparation

The twisted-bilayer $WSe_2$ and few-layer 2H-$WSe_2$ samples were fabricated by the mechanical exfoliation of bulk 2H-$WSe_2$ (2D Semiconductors) and dry-transfer method using Elvacite 2552C (Lucite International Inc.) as a polymer stamp [50]. The twist angles were set to 0.3° by tear-and-stack method [51], although the actual twist angle slightly varies around this value depending on the position. The $WSe_2$ flakes were sandwiched between a few layers of hexagonal boron nitride (hBN) and then placed on an amorphous silicon nitride ($SiN_x$) substrate (Norcada Inc.). We prepared two twisted-bilayer $WSe_2$ samples (Sample #1 and #2), which are used for temperature-dependent and time-resolved measurements, respectively. We use 20-nm and 100-nm $SiN_x$ substrate for Sample #1 and #2, respectively. The local twist angle $\theta$ of twisted-bilayer $WSe_2$ is determined from the domain size $L$ using the relation $L = a/[2\sin(\theta/2)]$, where $a$ = 3.32 Å is lattice constant.

**Drift correction of time-dependent data**

For the time-dependent dark-field image measurement in Fig. 4, we performed time-dependent dark-field image measurements repeatedly, and these data are summed to improve the signal-to-noise ratio. Before the summation, the sample drift between several measurements was corrected. We numerically evaluated the cross-correlation of dark-field images as follows:

$$A_{t,n}(x_i, y_j) = \sum_{p,q=-N/4}^{N/4} I_{t,n}(x_p, y_q) I_{\text{ref}}(x_p - x_i, y_q - y_j),$$

where $I_{t,n}(x_p, y_q)$ is the dark-field image at delay time $t$ of $n$-th measurement, $I_{\text{ref}}(x_p, y_q)$ is the dark-field image at $t$ = -100 ps of the first measurement, $x_{i,p}, y_{j,q}$ represents the pixelated position in the image, $N$ = 256 is the image size. Therefore, $A_{t,n}(x_i, y_j)$ is the cross-correlation of the central part of the image. We estimated the shift of the dark-field image as the maximum position of $A_{t,n}(x_i, y_j)$, by which the respective image was shifted. The typical magnitude of the shift was 0-4 pixels. All analyses were done using multi-dimensional data analysis and visualization platform lys [52].

**Density-functional calculations**

Density-functional theory (DFT) calculations using VASP package [53] were performed to evaluate the structural parameters of AB and 2H stacking bilayer $WSe_2$ and bulk 2H-$WSe_2$. It is noted that we do not include the effect of twisted crystal structure for AB stacking bilayer $WSe_2$. The calculations were performed in the framework of the van der Waals density functional theory [54] using the pseudo-potential method in conjunction with the generalized gradient approximation. The wave functions were expanded in the plane wave basis with the cutoff energy of 500 eV to ensure good convergence of the stress and elastic tensor. The *k*-point sampling was set to 9 × 9 × 4 in the first Brillouin zone. The a axis length was fixed to 3.32 Å, which is experimentally determined by the electron diffraction measurement. The interlayer distance and selenium height were relaxed with a force convergence of 0.0015 eV/Å. Although the two selenium heights are relaxed independently, the difference between them are less than 0.001 Å.

**Acknowledgments:**
This work was supported by JSPS KAKENHI (Grant numbers 25K00057, JP20H01834, JP21H05235 , JP24K01285, 24H00410, JP25H01251), JST PRESTO (Grant number JPMJPR24JA), CREST (Grant number. JPMJCR20B4). T. S. acknowledges support from Toray Science Foundation (#23-6405)、The Sumitomo Foundation. K.W. and T.T. acknowledge support from the JSPS KAKENHI (Grant numbers 21H05233 and 23H02052), JST CREST (JPMJCR24A5), and World Premier International Research Center Initiative (WPI), MEXT, Japan.